\documentclass[a4paper,12pt]{article}
\usepackage{mathptmx}
\usepackage{epsfig}
\usepackage{float}
\usepackage{latexsym}
\usepackage{amsfonts}
\newcommand{\beq}{\begin{equation}}
\newcommand{\eeq}{\end{equation}}
\newcommand{\beqa}{\begin{eqnarray}}
\newcommand{\eeqa}{\end{eqnarray}}
\newcommand{\nn}{\nonumber \\}
\newtheorem{definition}{\mbox{Definition}}
\newtheorem{lemma}{\mbox{Lemma}}
\newtheorem{remark}{\mbox{Remark}}
\newtheorem{theorem}{\mbox{Theorem}}
\def \h {\frac{1}{2}}
\def \hp {\hat{p}}
\def \li {\mathrm {Li}}
\def \r {\rho}
\def \t {\tau}
\def \D {D}
\def \N {\mathbb N}
\def \P {\mathbb P}
\def \R {\mathbb R}
\begin{document}
\title{\bf Prime Number Diffeomorphisms, Diophantine Equations and
the Riemann Hypothesis}
\author{Lubomir Alexandrov and Lachezar Georgiev \\ \\
{\it Institute for Nuclear Research and Nuclear Energy }\\
{\it Tsarigradsko Chaussee 72, 1784 Sofia, Bulgaria} \\
{\it and} \\
{\it Bogoliubov Laboratory of Theoretical Physics, JINR,} \\
{\it 141980 Dubna, Russia} \\
 \\
}

\maketitle
\begin{abstract}
We explicitly construct a diffeomorphic pair $(p(x),p^{-1}(x))$ in
terms of an appropriate  quadric spline interpolating the prime
series. These continuously differentiable functions are the smooth
analogs of the prime series and the prime counting function,
respectively, and  contain the basic information about the
specific behavior of the primes. We employ $p^{-1}(x)$ to find
approximate solutions of Diophantine equations over the primes and
discuss how this function could eventually be used to analyze the
von Koch estimate for the error in the prime number theorem which
is known to be equivalent to the Riemann hypothesis.
\end{abstract}
\mbox{\quad}
\section{Introduction}
We shall use the following notation: $\N$ is the natural numbers
set,  $p_n$ or $p(n)$ is the $n$-th prime, $\hat{p}_n=p_{n+1}-
p_{n} -1$ is the number of composites in the interval $(p_{n},
p_{n+1})$, $\P$ is the prime numbers set, $\pi(x)$ is the prime
counting function, $\li(x)$ is the logarithmic integral , i.e.,
\beq\label{Li} \N=\{1,2,3,\ldots \}, \quad \P=\{p(n): \ n\in\N \},
\quad \pi(x)=\sum_{p\leq x , \ p\in \P} 1,\quad
\li(x)=\int\limits_{2}^{x}\frac{ds}{\ln(s)}. \eeq The main
objective of this paper is to show that the following pair of
one-to-one mappings
\[
p(n) : \N \to \P, \quad p^{-1}(q): \P\to \N
\]
could be extended to a pair of diffeomorphisms over the real semi-axis
$(0,\infty)$.
\begin{definition}\label{def:diff}
The pair of functions $f(x)\in C^{(1)}(0,\infty)$,  $g(x)\in C^{(1)}(1,\infty)$
is called  {\bf prime number diffeomorphic} if the following
conditions hold
\beqa
f(n)&=&p_n, \quad n\in \N \label{f_n} \\
f\left(n+\frac{1}{2}\right)&=&\frac{1}{2}\left( p_n+p_{n+1}\right), \quad n\in\N
\label{f_n2}\\
f(g(x))&=&x \quad \forall x\in (1,\infty), \label{f-g}\\
g(f(x))&=&x \quad \forall x\in (2,\infty), \label{g-f}\\
\pi(x)&=&\lfloor g(x) \rfloor  \label{pi_x}.
\eeqa
The diffeomorphisms $f(x)$ and $g(x)$ are called {\bf the prime curve} and
the {\bf prime counting curve}, respectively.
\end{definition}
The function $\pi_R(x)$ of Riemann--Von Mangoldt (\cite{edw}, p. 34, (2), (3))
which can be expressed in terms of the zeros of the Riemann zeta function
 is the closest, among all known function, to the prime counting curve.
However, the function $\pi_R(x)$ is not invertible and cannot be used
for the correspondence  to the appropriate  prime curve.
It turns out that one can take the opposite way: construct an invertible
interpolation of the prime series and then obtain from it a smooth counting
curve. In this paper we describe such an invertible interpolant which
is found among differentiable polynomial splines of minimal degree.

These diffeomorphisms allows us to use in a natural way Fourier
analysis as well as iterative methods for the solution of
nonlinear problems in the case when it is necessary to account for
the specific character of the non asymptotic
 behavior of  the primes.
As an example of the application of the above diffeomorphisms we
consider in Sect.~\ref{sec:diophant} an approximate method for the
solution of Diophantine
 equations over $\P$.

We shall prove in Sect.~\ref{sec:RH} that the differentiable
function $p^{-1}(x)$ has the same
asymptotic behavior as $\pi(x)$ for $x\to\infty$. This could be used
for the analysis of the validity of the von Koch estimate in the form
$|p^{-1}(x)-\li(x)|/\sqrt{x}\ln(x)\sim \mathrm{const.}$,
which is known to be equivalent to the Riemann hypothesis (RH) \cite{edw,pom}.
\section{Prime number diffeomorphisms based on a quadric spline}
\label{sec:diff}
Let us define the following functions
\beqa
a^-_n(x)&=&-2\hat{p}_{n-1} (x-n)^2 + (x-n) +p_n, \quad x>0,\quad
n=2,3, \ldots,\nn
a^+_n(x)&=&2\hat{p}_n \left(x-n-\h\right)^2 +
(2\hp_n+1)\left(x-n-\h\right)+ \frac{p_n+p_{n+1}}{2}. \nonumber
\eeqa
Their derivatives are
\[
\frac{d a^-_n(x)}{dx}= 4 \hat{p}_{n-1}(n-x)+1, \quad
\frac{d
a^+_n(x)}{dx}= 4 \hat{p}_{n}(x-n)+1.
\]
For any $n=2,3,\ldots$ the  functions are sewed together
\beqa
a^-_n(n)&=&a^+_n(n)=p_n,\label{a_nn}\\
\left.\frac{d a^-_n(x)}{dx}\right|_{x=n}&=& \left.\frac{d
a^+_n(x)}{dx}\right|_{x=n}=1,\label{da_nn}\\
a^-_{n+1}\left(n+\frac{1}{2}\right)&=&a^+_n\left(n+\frac{1}{2}\right)=
\frac{1}{2}(p_n+p_{n+1}),\label{a_nn2}\\
\left.\frac{d
a^-_{n+1}(x)}{dx}\right|_{x=n+\frac{1}{2}}&=&\left.\frac{d
a^+_n(x)}{dx}\right|_{x=n+\frac{1}{2}} =2\hp_n+1 .\label{da_nn2}
\eeqa Equations (\ref{a_nn}), (\ref{da_nn}), (\ref{a_nn2}) and
(\ref{da_nn2}) define the following continuously differentiable
quadric spline \beq\label{p} p(x)=\left\{\begin{array}{lll}
x+1, & 0<x\leq \frac{3}{2}, & \\
a^-_n(x), & n-\h \leq x \leq n, & n=2,3,\ldots , \\
a^+_n(x), & n \leq x \leq n+\h, & n=2,3, \ldots \;. \\
\end{array} \right.
\eeq
with first derivative
\beq\label{dpdxdef}
\frac{d p(x)}{d x}=\left\{\begin{array}{lll}
1, & 0<x\leq \frac{3}{2}, & \\
4 \hat{p}_{n-1}(n-x)+1, & n-\h\leq x \leq n, & n=2,3,\ldots , \\
4 \hat{p}_{n}(x- n)+1, & n \leq x \leq n+\h, & n=2,3, \ldots \; .\\
\end{array} \right. .
\eeq
Inverting the function $a^-_n(x)$  in the interval $n-\h\leq x \leq n$
and $a^+_n(x)$ in the interval $n\leq x \leq n+\h$
gives the following inverse functions and their derivatives
\beqa
b^-_n(x)&=&n+\frac{1-\left(8\hp_{n-1}(p_n-x)+1 \right)^{\h}}{4\hp_{n-1}}, \nn
b^+_n(x)&=&n+\frac{\left(8\hp_n(x-p_n)+1 \right)^{\h}-1}{4\hp_n},\nn
\frac{d b^-_n(x)}{d x}&=& \left(8\hp_{n-1}(p_n-x)+1  \right)^{-\h} , \nn
\frac{d b^+_n(x)}{d x} &=& \left(8\hp_n(x-p_n)+1 \right)^{-\h}. \nonumber
\eeqa
The functions $b^-_n(x)$,  $b^+_n(x)$ and their derivatives are sewed
together in a similar way like Eqs.~(\ref{a_nn}), (\ref{da_nn}),
(\ref{a_nn2}) and (\ref{da_nn2}):
\beqa
b^-_n(p_n)&=&b^+_n(p_n)=n,\label{b_nn}\\
\left.\frac{d b^-_n(x)}{dx}\right|_{x=p_n}&=&
\left.\frac{d b^+_n(x)}{dx}\right|_{x=p_n}=1,\label{db_nn}\\
b^-_{n+1}\left(\frac{p_n+p_{n+1}}{2}\right)&=&
b^+_n\left(\frac{p_n+p_{n+1}}{2}\right)=
n+\h,\label{b_nn2}\\
\left.\frac{d b^-_{n+1}(x)}{dx}\right|_{x=\frac{p_n+p_{n+1}}{2}}&=&
\left.\frac{d b^+_{n}(x)}{dx}\right|_{x=\frac{p_n+p_{n+1}}{2}}=
\frac{1}{2\hp_n+1}.\label{db_nn2}
\eeqa
Finally Eqs. (\ref{b_nn}), (\ref{db_nn}), (\ref{b_nn2}) and (\ref{db_nn2})
define the continuously differentiable  inverse spline
\beq\label{p_}
p^{-1}(x)=\left\{\begin{array}{lll}
x-1, & 1 <x\leq \frac{5}{2}, & \\
n+\frac{1-\left(8\hp_{n-1}(p_n-x)+1  \right)^{\h}}{4\hp_{n-1}}, &
 \frac{p_{n-1}+p_{n}}{2}  \leq x \leq p_n, & n=2,3, \ldots , \\
n+\frac{\left(8\hp_n(x-p_n)+1  \right)^{\h}-1}{4\hp_n}, & p_n\leq
x \leq \frac{p_{n}+p_{n+1}}{2}, & n=2,3,\ldots ,
\end{array} \right.
\eeq
with first derivative
\beq\label{dp-1dxdef}
\frac{d p^{-1}(x)}{d x}=
\left\{\begin{array}{lll}
1, & 1 <x\leq \frac{5}{2}, & \\
\left(8\hp_{n-1}(p_n-x)+1  \right)^{-\h}, &
 \frac{p_{n-1}+p_{n}}{2}  \leq x \leq p_n, & n=2,3, \ldots ,\\
\left(8\hp_n(x-p_n)+1  \right)^{-\h}, & p_n \leq x \leq
\frac{p_{n}+p_{n+1}}{2} , & n=2,3,\ldots\; .
\end{array} \right. .
\eeq

\begin{figure}[H]
\begin{center}
\epsfig{file=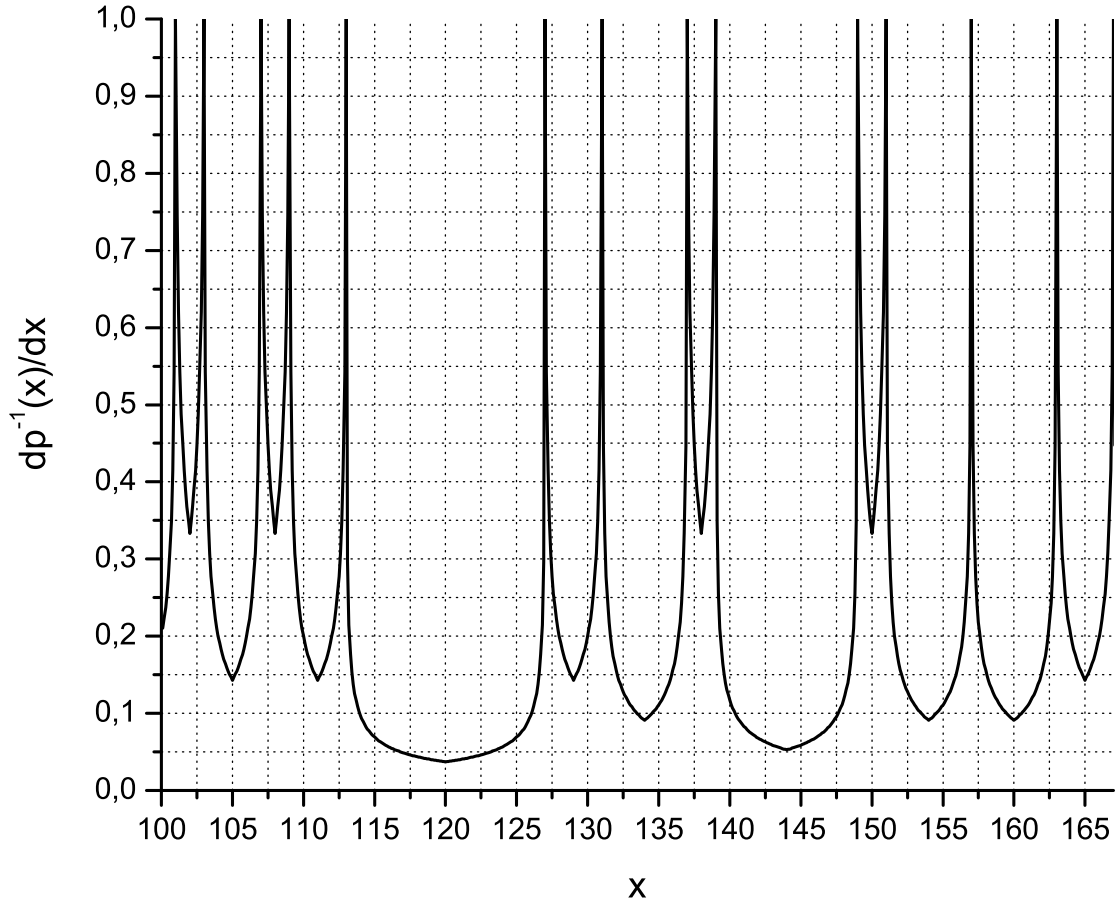,height=13cm,clip=,%
bbllx=30,bblly=30,bburx=514,bbury=431}
\caption{The derivative of $p^{-1}(x)$ \label{fig:dp-1dx}}
\end{center}
\end{figure}

\begin{lemma}The derivatives of $p(x)$ and $p^{-1}(x)$ satisfy
the following inequalities
\beqa
&1 \leq \frac{d p(x)}{d x} < \infty ,\quad  x>0,& \label{dpdx}\\
&0 < \frac{d p^{-1}(x)}{d x} \leq 1 , \quad x>1. & \label{dp-1dx}
\eeqa
\end{lemma}
\textit{Proof:}
 The above inequalities follow directly from the definitions
(\ref{dpdxdef}) and  (\ref{dp-1dxdef}). \\
 \noindent
 $\Box$\\
\noindent
In the rest of this section we shall prove the following
\begin{theorem}\label{thm:1}
\noindent
\begin{description}
\item[(i)]{The pair $(p(x),p^{-1}(x))$ is  prime number diffeomorphic. }
\item[(ii)]{The specific behavior of the prime and counting curves are
traced by the invariants:} \beqa 1&=&\left. \frac{d
p(x)}{dx}\right|_{x=n}  = \left. \frac{d p^{-1}(x)}{d
x}\right|_{x=p_n}, \quad n=2,3,\ldots,
\label{dp-1dxpn}\\
-1&=&\mathrm{sign}\left( \left. \frac{d^2 p(x)}{d
x^2}\right|_{x=n-0} \right) \mathrm{sign}\left( \left. \frac{d^2
p(x)}{d x^2}\right|_{x=n+0} \right) , \quad
n=3,4, \ldots, \label{d2pdx}\\
-1&=&\mathrm{sign}\left( \left. \frac{d^2 p^{-1}(x)}{d
x^2}\right|_{x=p_n-0} \right) \mathrm{sign}\left( \left. \frac{d^2
p^{-1}(x)}{d x^2}\right|_{x=p_n+0} \right) , \quad n=3,4,\ldots\;.
\label{d2p-1dx}
\eeqa
\end{description}
\end{theorem}
\textit{Proof}
\begin{description}
\item[(i)]{\quad \ According to the definitions (\ref{p}), (\ref{dpdxdef}),
(\ref{p_}) and (\ref{dp-1dxdef})  we have
the inclusions
$p(x)\in C^{(1)}(0,\infty)$ and $p^{-1}(x)\in C^{(1)}(1,\infty)$.
The validity of the interpolation conditions (\ref{f_n}) and (\ref{f_n2})
follows from Eqs.~(\ref{a_nn}) and (\ref{a_nn2}). The mutual
invertibility of $p(x)$ and $p^{-1}(x)$ follows from the fact that these
functions are continuous and monotonically increasing (see (\ref{dpdx}) and
(\ref{dp-1dx})) and the conditions (\ref{f-g}) and (\ref{g-f}) can be checked
directly. Equations~(\ref{b_nn}) show that  the functions $\pi(x)$ and
$p^{-1}(x)$ are related by the identity (\ref{pi_x}).
}
\item[(ii)]{\quad \
Equations~(\ref{dp-1dxpn}) follow from Eqs.~(\ref{da_nn}) and (\ref{db_nn}).
The discontinuity of the second derivatives
\[
\frac{d^2p(x)}{dx^2}=\left\{\begin{array}{rl}
0, & 0< x\leq \frac{3}{2} \; , \\
-4\hp_{n-1}, & n-\h\leq x \leq n , \quad n=2,3,\ldots \;, \\
4\hp_n, & n\leq x \leq n+\h , \quad n=2,3,\ldots \;, \\
  \end{array}\right.
\]
and
\[
\frac{d^2p^{-1}(x)}{dx^2}=\left\{\begin{array}{ll}
0, & 0< x\leq \frac{5}{2} , \\
4\hp_{n-1} (8\hp_{n-1}(p_n-x)+1)^{-3/2}, &
\frac{p_{n-1}+p_n}{2} \leq x \leq p_n \quad , n=2,3,\ldots, \\
-4\hp_{n}(8\hp_{n}(x-p_n)+1)^{-3/2}, & p_n \leq x \leq
\frac{p_{n}+p_{n+1}}{2} \quad , n=2,3,\ldots, \\
  \end{array}\right.
\]
implies Eqs.~(\ref{d2pdx}) and (\ref{d2p-1dx}). This completes the
proof.\\
\noindent $\Box$ }
\end{description}
\begin{remark}
The fact that the coefficients of $p(x)$ are integers as well as
its invertibility follow from Eq.~(\ref{f_n2}) in
Definition~\ref{def:diff} (i.e., from Eqs.~(\ref{a_nn2})).
\end{remark}
\begin{figure}[H]
\begin{center}
\epsfig{file=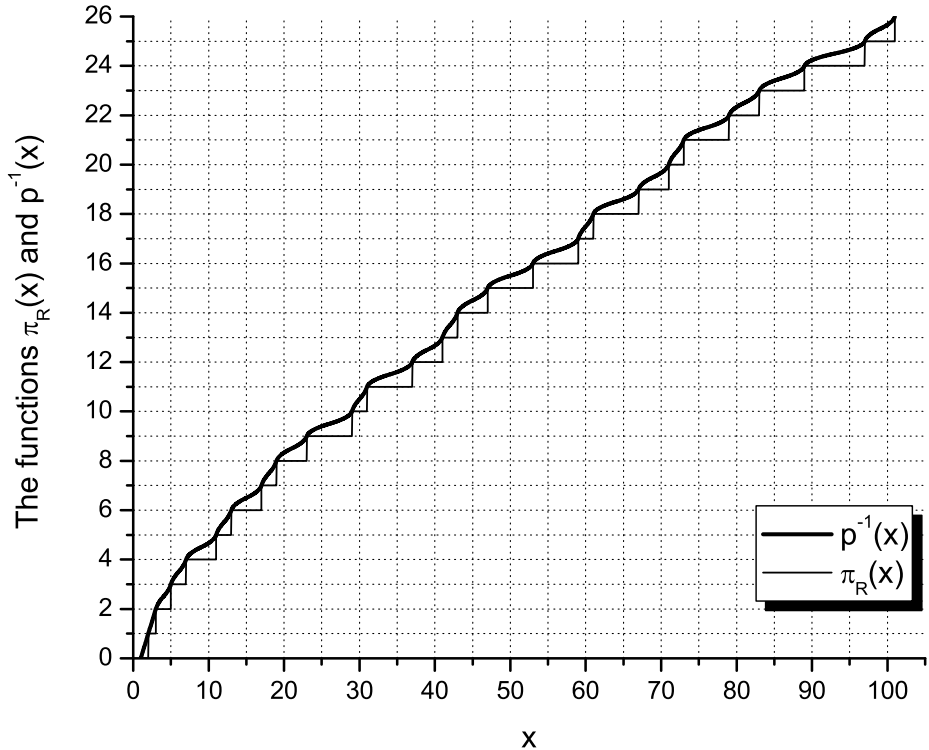,width=13cm,clip=,%
bbllx=30,bblly=30,bburx=512,bbury=432}
\caption{The Riemann--von Mangoldt counting step function $\pi_R(x)$ and
its continuously differentiable analog  $p^{-1}(x)$.
\label{fig:p-1pi}}
\end{center}
\end{figure}
The comparative plot of the functions $p^{-1}(x)$ and $\pi_R(x)$ is
shown on Fig.~\ref{fig:p-1pi}.
The derivative  $dp^{-1}/dx$ is  shown on  Fig.~\ref{fig:dp-1dx}.
Its oscillating nature as well as the fact that it takes values
between $0$ and $1$ is obvious from this figure.
The diffeomorphisms $p(x)$, $p^{-1}(x)$ and their derivatives are
realized in a Fortran90 program package
called {\tt pp\verb"_".f90} which can be found at this
URL \cite{p_.f90}.
\section{Approximate solution of Diophantine equations}
\label{sec:diophant}
\begin{definition}
Given a set of strictly increasing functions $h_i(x)\in C^{(1)}(0,\infty)$, $i=1,\ldots n$   the system
\beqa
&&f_1(x_1,\ldots,x_n)=0, \label{sys1}\\
&&f_2(h_1(x_1),\ldots,h_n(x_n)):=\sum\limits_{i=1}^n\sin^2(\pi h_i(x_i))=0 ,
\label{sys2}
\eeqa
where Eq. (\ref{sys1}) is a Diophantine one
is called real-Diophantine on the real semi axis $(0,\infty)$.
\end{definition}
The real-Diophantine systems allow us to find solutions of Diophantine
equations in terms of real approximations
of integer numbers. For this purpose one should apply numerical methods which
work even when the derivative is degenerate at the solution (see, e.g.,
\cite{A2} and \cite{kalt}).

Let us write the system (\ref{sys1}), (\ref{sys2}) in a vector
form as follows
\beq\label{vector}
Fx=0,
\eeq
where
\[
Fx={f'}^T(x) fx, \quad fx=\left[ f_1(x),f_2(x)\right]^T, \quad
f:\D_f \subset \R^n\to\R^m, \quad x\in\R^n,\quad n>m,
\]
$f'(x)$ is the Jacobi matrix and $\D_f $ is an open
convex domain in $\R^n$.

Here we shall quote an autoregularized version of the
Gauss--Newton method \cite{A1,A3} as one of the possible methods
for the solution of Eq.~(\ref{vector}): \beqa\label{GN} x^0\in
\D_f, \quad \epsilon_0>0, \quad &&\left( {f'}^T(x^k) f'(x^k)
+\epsilon_k I \right)(x^{k+1}-x^k)=-Fx^k,
\quad k=1,2,\ldots\; . \\
&&\epsilon_k=\h\left( \sqrt{\t_k^2+4c\r_k} -\t_k\right), \nn
&&\t_k=|| {f'}^T(x^k) f'(x^k)   ||_\infty, \quad \r_k=
||Fx^k||_\infty, \quad
c=\frac{\epsilon_0+\epsilon_0\t_0}{\r_0},\nonumber \eeqa where $||
\cdot ||_\infty$ is the uniform vector or matrix norm, and the SVD
method \cite{golub} is assumed for the solution of the linear
problem in (\ref{GN}). In order to find all solutions of
Eq.~(\ref{vector}) in the domain $\D_f $ the vector $Fx^k$ is
repeatedly multiplied by the local \textbf{root extractor}
\[
e_j(x,\bar{x}^{(j)})= \frac{1}{1-\exp\left( - ||x - \bar{x}^{(j)}
||_2^2 \right)},
\]
in which $\bar{x}^{(j)} $ is the $j$-th solution of
Eq.~(\ref{vector}). In the repeated solutions of the transformed
problem
\beq\label{F_M}
F_{J} \,  x:=\left( \prod_{j=1}^J
e_j(x,\bar{x}^{(j)}) \right) Fx=0, \quad J\geq 1,
\eeq
the process
(\ref{GN}) is executed with a new $Fx:=F_J x$. For every solution
the process (\ref{GN}) is started many times with different $x^0$
and $\epsilon_0$. Each time when $J$ increases the derivatives
$f'(x_k)$ are computed analytically and the matrices
${f'}^T(x_k)f'(x_k)$  are adaptively scaled \cite{more}.
\textit{The necessary last  step of the method consists in
 a direct substitution check whether the Diophantine
equation residual vanishes exactly when the found solutions are rounded
to integers}.

The system (\ref{sys1}), (\ref{sys2}) has been considered in
Refs.~\cite{costa} and \cite{smale} (pp. 285--286) as
\textit{undecidable}. In Table~\ref{tab}  we present some examples
in which the real-Diophantine system (\ref{sys1}), (\ref{sys2}) is
solvable including the case when it is solvable over the primes
(see cases 2, 3, 4 and 6).
\begin{table}[htb]
\centering
\caption{Real-Diophantine systems: when $h_i(x_i)=x_i$ the solution is looked
for among the integers while $h_i(x_i)=p^{-1}(x_i)$ means that the solution
would be among the primes.\label{tab} }
\begin{tabular}{|c||c|c|c|c|c|}
\hline
case & $f_1(x)$ & $n$ &$m$& $h_i(x_i)$ & source \\
\hline \hline
1 & $x_1^2+ x_2^2 -x_3^2=0$  & 3 & 2 & $x_i$ & Pythagoras   \\
\hline
2 & $x_1^2+ x_2^2 -x_3^2 -1 =0$ & 3 & 2 & $p^{-1}(x_i)$ & Sierpinski  \\
\hline
3 & $x_1^2+ x_2^2 +x_3^2 + x_4^2 =n, \ n\in\N$ & 4 & 2 & $p^{-1}(x_i)$ & Lagrange  \\
\hline 4 & $\sum\limits_{i=1}^9 x_i^3 =n, \ n\in\N$ & 9 & 2 &
$x_i, \ p^{-1}(x_i)$ & Waring,
Khinchin  \\
\hline 5 & $\sum\limits_{i=1}^{19} x_i^4 =n, \ n\in\N$ & 19 & 2 &
$x_i$ & Waring  \\
\hline 6 & $ \left( x_1/x_2\right)^2-\left( x_3/x_4\right)^3
-x_5=0$ &  5 & 2 &
$h_i(x_i)=p^{-1}(x_i)$,  & Fermat--Bache \\
& & & &$i=1,\ldots,4$,\quad $h_5(x_5)=x_5 $ & \\
 \hline
\end{tabular}
\end{table}
Here we shall describe in more detail two special examples which emphasize
the crucial role of the last step of the above method. In the first
one we represent the prime number $5081$ as a sum of $9$ cubes of primes
(case $4$ in Table~\ref{tab} with $h_i(x_i)=p^{-1}(x_i)$).
We find  two different  solutions:
\beq\label{5081}
5081=  \left\{ \begin{array}{l}
 2\times 2^3 + 3 \times 3^3 + 5^3 + 2\times 11^3 + 13^3\;, \\
 3^3 + 3\times 5^3 + 2\times 7^3 + 3\times 11^3\;.
\end{array}\right.
\eeq
Notice that this problem has been solved as a
real-Diophantine system on a machine with $16$ significant
figures. The unknowns $2, \ 3, \ 5, \ 7, \ 11$ and $13$
in the first line of Eq.~(\ref{5081}) have been found
with $9$ significant figures at residual
$||f_1(x)||_\infty=10^{-14}$. A convergent process of the kind
(\ref{GN})  has been build after $36$ unsuccessful attempts which
costed $5634$ iterations with $4$ different initial guesses $x_0$
combined with $9$ different initial regularizators $\epsilon_0$.
The last step of the method yields  an exact equality in
Eq.~(\ref{5081}). It would be interesting to investigate
whether the number  of primes which can be represented as the
sum of $9$ prime cubes is infinite.

In the second example we consider the equation
\beq\label{FB}
\left( \frac{x_1}{x_2}\right)^2 - \left( \frac{x_3}{x_4}\right)^3 =x_5,
\eeq
where $x_i$, $i=1,2,3,4$ are sought as primes while $x_5$ as integer
(special case of the unsolved  Fermat--Bache problem in which the solutions
are a rational pair   $(x_1/x_2,\ x_3/x_4)$ and an integer $x_5$).
The approximate solution found under conditions similar to those in
the previous example, Eq.~(\ref{5081}),
\beqa
&&x_1=787.000011, \quad\quad  x_2=348.99999357, \quad x_3=457.00002128, \nn
&&x_4=1049.0000001,\quad x_5=5.0024058062, \nonumber
\eeqa
leads to a nonzero residual after rounding to integers
\[
\left( \frac{787}{349}\right)^2 - \left( \frac{457}{1049}\right)^3 - 5=
0.002405488240.
\]
This example is an illustration of the crucial importance of the
last step of the method--the vector $(787,\ 349, \ 457, \ 1049, \
5)$ is not a true solution of the Fermat--Bache equation
(\ref{FB}).

The above method for the solution of Diophantine equations
works because of a combination of factors:
autoregularization, SVD method, adaptive scaling, and because
the solutions of Diophantine equations are well isolated.
There is a semi-local convergence theory
\cite{A1} in the non-degenerate case however no justification
in the degenerate case is available by now.
The methods of Refs.~\cite{A2,kalt} are applied to this
problem with little success.

\begin{remark}
The method (\ref{sys1}), (\ref{sys2}) for the solution of the
Diophantine equation (\ref{sys1}) does not contradict the negative
solution of the 10-th Hilbert's problem \cite{mat} because our solutions
 are approximated  in a bounded domain.
\end{remark}
\section{The function $p^{-1}(x)$ and the  Riemann hypothesis}
\label{sec:RH}
\begin{lemma}\label{lemma}
The functions $p^{-1}(x)$ and $\pi(x)$ are related by
\[
|p^{-1}(x)-\pi(x)|\leq 1\quad \forall x>1.
\]
\end{lemma}
\textit{Proof:} Let us assume that $p_n\leq x \leq p_{n+1}$ for some $n$.
According to Eq.~(\ref{dp-1dx}) the function $p^{-1}(x)$
is strictly increasing, i.e., $p^{-1}(x)\leq p^{-1}(p_{n+1})$. On
the other hand $\pi(p_{n}) \leq\pi(x)\leq \pi(p_{n+1})$ so that
\[
|p^{-1}(x)-\pi(x)|\leq |p^{-1}(p_{n+1})-\pi(p_n)|=|n+1-n|=1,
\]
where we have used that  $p^{-1}(p_n)=n$ which follows from
Theorem~\ref{thm:1} and  Eq.~(\ref{g-f}).

\noindent
$\Box$

\begin{theorem}
The asymptotics of $p^{-1}(x)$ is the same as
that for $\pi(x)$ when $x\to\infty$.
\end{theorem}
\textit{Proof:}
Let us consider the relative difference  of  $\pi(x)$ and $\li(x)$.
According to Lemma~\ref{lemma} we can write
\[
\frac{|\pi(x)-\li(x)|}{\li(x)}\leq \frac{|\pi(x)-p^{-1}(x)|}{\li(x)}+
\frac{|p^{-1}(x)-\li(x)|}{\li(x)}\leq  \frac{1}{\li(x)}+
\frac{|p^{-1}(x)-\li(x)|}{\li(x)}  .
\]
Therefore in the limit $x\to\infty$  we can ignore the term $1/\li(x)$ and
investigate $p^{-1}(x)$ instead of  $\pi(x)$.

\noindent
$\Box$

\subsection{Differential equation and the von Koch estimate}
Let us consider the following function
\beq\label{K}
K(x)=\frac{p^{-1}(x)-\li(x)}{\sqrt{x}\, \ln(x)}, \quad x>1,
\eeq
where $\li(x)$ is defined in Eq.~(\ref{Li}).
According to the von Koch estimate (see \cite{edw}, pp. 90)
the Riemann hypothesis is
equivalent to the statement that $K(x)$ is asymptotically constant, i.e.,
\[
\lim_{x\to \infty}K(x)=\mathrm{const.} \quad \Longleftrightarrow \quad
\mathrm{RH}.
\]
Because the function $p^{-1}(x)$ is continuously differentiable we
can write the following differential equation for $K$
\beq\label{K'}
K'(x)=-\left(\frac{1}{2x}+ \frac{1}{x\, \ln(x)}
\right)K(x)+ \frac{\frac{dp^{-1}(x)}{dx}
-\frac{1}{\ln(x)}}{\sqrt{x} \, \ln(x)}.
\eeq
The derivative
$dp^{-1}(x)/dx$ is strongly oscillating as shown on
Fig.~\ref{fig:dp-1dx}, however it is restricted between $0$ and
$1$ according to Eq.~(\ref{dp-1dx}). Therefore we shall consider
the solution of Eq.~(\ref{K'}) in the interval $p_n\leq x \leq
(p_n+p_{n+1})/2$, $n\to\infty$,  and shall use the fact that
(see Eqs.~(\ref{db_nn}) and (\ref{db_nn2}))
\beq\label{dp-1dxpn2}
\left. \frac{dp^{-1}(x)}{dx}\right|_{x=p_n}
=1, \quad \left. \frac{dp^{-1}(x)}{dx}\right|_{x=(p_n+p_{n+1})/2}
=\frac{1}{2\hp_n+1}.
\eeq
Thus, for  $x\to p_n$ we can substitute
$dp^{-1}/dx=1$ in Eq.~(\ref{K'}), neglect the term $1/\ln(x)$ in
the  limit $x\to\infty$ and solve the equation
\beq\label{KI}
K'(x)=-\left(\frac{1}{2x}+ \frac{1}{x\, \ln(x)} \right)K(x)+
\frac{1}{\sqrt{x} \, \ln(x)}. \eeq The general solution of this
equation can be written as
\[
K(x)=\frac{\sqrt{x}}{\ln(x)} +\frac{{c'}_n}{\sqrt{x}\, \ln(x)},
\]
where the first term in the right-hand-side is a partial solution of the
inhomogeneous equation (\ref{KI}) while the second one is the general
solution of the homogeneous equation and $c'_n$ is a constant.

At the right-hand border $x\to (p_n+p_{n+1})/2$,
we can neglect,  for $x\to \infty$, the term $dp^{-1}/dx=(2\hp­_n+1)^{-1}$
and keep only
$1/\ln(x)$ assuming that $\hp_n\sim p_n^\alpha$ with $\alpha>0$.
In this case we should solve the equation
\beq\label{KII}
K'(x)=-\left(\frac{1}{2x}+ \frac{1}{x\, \ln(x)} \right)K(x) +
\frac{-1}{\sqrt{x} \, \ln^2(x)}.
\eeq
The general solution of (\ref{KII}) is again the sum of a partial
solution (the first term bellow) of the inhomogeneous equation and
the general solution  (the second term) of the homogeneous one
\[
K(x)=-\frac{\li(x)}{\sqrt{x}\, \ln(x)} +\frac{{c''}_n}{\sqrt{x}\, \ln(x)}.
\]
Substituting the logarithmic integral with its leading term
for $x\to\infty$, i.e.,  $\li(x)\sim x/\ln(x)$ we can finally write
\[
K(x)\sim\left\{ \begin{array}{rl}
\frac{\sqrt{x}}{\ln(x)} +\frac{c'_n}{\sqrt{x}\, \ln(x)}, & x\to p_n \\
-\frac{\sqrt{x}}{\ln^2(x)} +\frac{{c''}_n}{\sqrt{x}\, \ln(x)}, & x\to
\frac{p_n+p_{n+1}}{2}
\end{array} \right.
\]
It is tempting to regard the general solution of the homogeneous equation
as subleading in the limit $x\to \infty$ and the first terms as
expressing the oscillations of $K(x)$ close to the borders of the
considered intervals. However, let us note that the constants $c'_n$
and ${c''}_n$
might depend on the primes gap $\hp_n$
which on its own depends on $p_n$ and this last dependence is currently
unknown.
\subsection{The l'Hospital rule}
Here we shall consider the limit
\[
\lim_{x\to \infty} \frac{p^{-1}(x)-\li(x)}{\li(x)}=
\lim_{x\to \infty} \frac{p^{-1}(x)}{\li(x)} -1
\]
Because the function $p^{-1}(x)$ is differentiable we can apply
the l'Hospital rule if the limit
\[
\lim_{x\to \infty} \frac{\frac{dp^{-1}(x)}{dx}}{\frac{1}{\ln(x)}} =
\lim_{x\to \infty} \ln(x)\frac{dp^{-1}(x)}{dx}
\]
exists. Now let us show that if the RH is true then this limit does
not exist. Indeed, let us choose two subsequences of $x\to\infty$, namely
\beqa
(i)&& \quad x=p_n, \qquad \qquad n\to\infty, \nn
(ii)&& \quad x=\frac{p_n+p_{n+1}}{2}, \quad n\to\infty .
\eeqa
Then, using again the values (\ref{dp-1dxpn2}) of the derivative
over (i) and (ii) we get
\beqa
(i)&& \quad \ln(x)\frac{dp^{-1}(x)}{dx} \sim \ln(p_n) \to \infty, \nn
(ii)&& \quad \ln(x)\frac{dp^{-1}(x)}{dx} \sim \frac{\ln(p_n)}{2\hp_n+1}\to 0.
\eeqa
The second limit follows from the statement that if the RH is true then
$\hp_n\sim p_n^{\h+\epsilon}$ for any $\epsilon>0$ when $n\to\infty$
\cite{pom}.
Most of the current  estimates of the primes gap $\hp_n$ lead to the
non-applicability of the l'Hospital rule. Nevertheless we cannot be sure until
a  rigorous estimate is found.


\section{Conclusions}
We have constructed a pair of diffeomorphisms $p(x)$ and
$p^{-1}(x)$ which interpolate the prime series and the prime
counting function, respectively, which are convenient for both
numerical and analytical applications. To the best of our
knowledge this is the first differentiable and invertible
interpolation of the prime series.

The function $p^{-1}(x)$  can be effectively used for
the solution of Diophantine equations which can be exploited
in many cases where the other methods do not work and could be
particularly useful when Diophantine equations are subsystems of
more complex real systems.

Because $p^{-1}(x)$ has the same behavior as $\pi(x)$ for $x\to\infty$
it could give more information about the asymptotic and non-asymptotic
distribution of primes. Perhaps, this could be used to draw some
conclusions about the Riemann hypothesis when more information about
the primes gaps becomes available.\\ \\

\noindent
\textbf{\Large Acknowledgments}\\ \\
The authors thank the BLTPh laboratory JINR, Dubna  for
hospitality and support. LG has been partially supported by the
FP5-EUCLID Network Program
 of the European Commission under Contract No. HPRN-CT-2002-00325
and by the Bulgarian National
Foundation for Scientific Research under Contract No. Ph-1406.



\begin{thebibliography}{999}

\bibitem{edw} H.M. Edwards, \textit{Riemann's Zeta Function},
Dover Publication, Mineola, New York (2001).

\bibitem{pom} R. Crandall, Carl Pomerance, \textit{Prime Numbers: A
Computational Perspective}, Springer, New York (2002).



\bibitem{p_.f90} L. Georgiev's homepage:
http://theo.inrne.bas.bg/$\sim$lgeorg/pp\verb"_".html



\bibitem{A2} L. Alexandrov, Regularized trajectories
for Newton kind approximations of the solutions of nonlinear equations,
\textit{Differential Equations}, {\bf XIII}, No. 7,  1281--1292
(1977, Russian).


\bibitem{kalt} B. Kaltenbacher, A. Neubauer, and A.G. Ramm,
Convergence of the continuous regularized Gauss--Newton method,
\textit{J. Inv. Ill-Posed Problems}, {\bf 10}, No. 3, 261--280 (2002).

\bibitem{A1} L. Alexandrov, Regularized
Newton--Kantorovich computational processes, \textit{J. Compt.
Math. Math. Phys.}, {\bf 11}, 36--43 (1971, Russian)

\bibitem{A3} L. Alexandrov,
\textit{Autoregularized Newton--Kantorovich iterational
processes}, JINR Commun. P5-5515, Dubna, 1970.


\bibitem{golub} G.H. Golub, C. Reinish, Singular Value
 Decomposition  and Least Squares, in \textit{Handbook for automatic
computation}, J.H. Wilkinson and C. Reinish, Eds., v. II, Linear Algebra,
Heidelberg, Springer, 1971.

\bibitem{more} J.J. More, The Levenberg--Marquardt algorithm, in
\textit{Numerical Analysis}, G.A. Watson, Ed., Lecture Notes in Math.,
630, Springer, Berlin 105--116 (1977).


\bibitem{costa} N.C.A. da Costa, and F.A. Doria, Undecidability
and incompleteness in Classical Mechanics,
\textit{Int. J. Theor. Phys.}, {\bf 30}, No. 8, 1041--1073 (1991).

\bibitem{smale} S. Smale, Mathematical Problems for the Next Century,
in \textit{Mathematics: Frontiers and Perspectives}, V. Arnold et.al.,
Eds., 271--294, AMS, Providence, USA (2000).

\bibitem{mat} Y. Matiyasevich, \textit{Hilbert's Tenth Problem}, The MIT Press,
Cambridge, Mass. (1993).

\end{thebibliography}
\end{document}